\documentstyle[aps,prl,eqsecnum,preprint,tighten]{revtex}
\begin{document} 
 \title{Invariant  structure of the hierarchy theory
of  fractional  quantum  Hall  states  with  spin}   
 \author{M.   Milovanovi\'{c}\thanks{Address after January 1: Physics
Department, Technion, Haifa 32000, Israel}    and   N.   Read}
\address{Departments  of Physics and Applied  Physics,  P.O.
Box  208284\\   Yale   University,   New  Haven,  CT  06520}
\date{\today} 
 \maketitle  
 \begin{abstract} 
We describe the invariant structure common to abelian  fractional quantum
Hall effect systems with spin. It appears in a generalization of the
lattice description of the polarized hierarchy that encompasses both
partially polarized and unpolarized ground state systems.  We formulate, 
using the spin-charge decomposition, conditions that should be
satisfied  so that the description
is SU(2) invariant. In the case of the spin-singlet hierarchy
construction, we find that there are as many SU(2) symmetries as there
are levels in the construction.  Various formalisms used before for hierarchies
(field-theoretic, algebraic, and wave-functions) are also used to show
the existence of a spin and charge lattice for the systems with spin.
The ``gluing'' of the charge and spin degrees of freedom in their bulk 
 is  described by the gluing theory of lattices.
The low-energy field theories and corresponding
quantum Hall lattices should serve as a starting point for the 
discussion of the stability of these systems.  
 \end{abstract}  
       \pacs{PACS:        }
\section{Introduction} 
 \label{introduction}  
Not long after
 Laughlin  proposed a theory \cite{laugh} for the  fractional  quantum  Hall
effect  (FQHE) at filling fraction $ \nu = 1/q,$ $\; q $ odd,  the  hierarchy  theory \cite{hald,halp} 
was  proposed  as an
explanation of the occurence of the FQHE at filling fraction
$p/q$  when  $p$ is  not  $1$  and  $q$  is, as in the
Laughlin  case, an odd integer.  It describes  new ground
states of these systems as  hierarchies of Laughlin  states of
quasiparticles;  at each level the source of  the quasiparticles
is the Laughlin  state of the previous level.  Some time later Jain
proposed  his  construction \cite{jain}  of some of these FQHE  states as filled
Landau  levels of  composite  fermions,  {\em  i.e.}  integer
quantum Hall states of   particles  that are  electrons
with an even number of flux quanta  attached.  We can view 
this  construction also as some kind of hierarchy and, indeed, in
\cite{read} it was proved that various constructions 
 (the standard  hierarchy  that we mentioned  first, the
Jain  construction) are different  descriptions of
a single   underlying    physical theory. This theory is described as
     a  lattice  of
excitations  \cite{read}, which contains all information  about the
quantum numbers and statistics of quasiparticles and properties of
the edge states.

If we introduce an  additional,  spin or pseudo-spin, degree of freedom, we
can construct two-component (ground) states, Halperin states
\cite{helv}  which are simple  generalizations  of the  Laughlin
state.  Some of the Halperin states that are spin-singlets   ({\em
i.e.} unique
 states with respect to the total spin) might  describe  the
systems  
for which $ \vec{S}_{\rm tot}^{2} $ and $ S_{\rm tot}^{z} $ are good quantum
numbers, and Zeeman energy is small \cite{fdm}.  A  hierarchy  of these  states, as an analog of the
standard  hierarchy in the completely  spin-polarized  case, was
first  described  in  \cite{rez}.  From the  field-theoretical
point of view, in a context of a very general  formulation  of
the FQHE with the spin degree of freedom, it was also  described  in
\cite{leeka}.  On the other hand the Halperin  states  contain a  very
simple  spin-charge  decomposition  \cite{more}, which  realizes
itself  on the edge, that  served as a  starting  point  for a
   hierarchy  in  \cite{lopez}.  Jain's   construction  of
spin-singlet  states \cite{jain} uses  Landau  levels  filled
with composite fermions of both spins.  Naturally, this can be
extended to  partially  polarized  states  with more  Landau
levels of one-spin particles than of the other \cite{jain,dudu}.
Some principles of the spin-singlet hierarchy construction were 
discussed in \cite{nawi} also.

The goal of our paper is to show the  existence
of a  basis-independent  description for  systems
with the spin  degree of  freedom,  no  matter  whether  their ground
states are
     partially     polarized, unpolarized or
spin-singlet.  But, in the  process  of  achieving  this, we use the
spin-charge decomposition and
find a structure of excitations  common to all of them, which
is interesting in its own right.

After a review of the Halperin states in Sec. II, in Sec. III  we
concentrate  on the hierarchy spin-singlet  systems. At the beginning,
the first-level hierarchy of a general, two-component system with
the (pseudo-)spin degree of freedom, is presented, based on the dual
bosonic Chern-Simons field theory. Then the spin-charge decomposition
is introduced, and the kind of ``gluing'' between the charge and spin
degrees of freedom of the excitations in the bulk of these systems is
described.   
This leads to information about the 
excitation  lattice    in this  case.  The
lattice  can be  easily  identified  as a  special  case  of
composite lattice  constructions  known in mathematics under the
name  ``gluing  theory'' \cite{glue}.  We  describe it   together  with  a
lattice,  which  lies  in  the   excitation   lattice,  that
corresponds to the order parameters of the systems. 
Also, in the context of the spin-charge decomposition, we use the
invariance under change of hierarchy basis to formulate the
spin-singlet condition.  One of the bases is identified as the basis of
Jain's construction.
The most important feature of the hierarchy of spin-singlet
states, which the lattices incorporate, is the existence of as many
independent  SU(2) symmetries as there are levels in the
hierarchy constructions.  Sec.
IV and Sec. V describe, in the new  formalism, a  generalized
spin-singlet   hierarchy  and  the  hierarchy  of  partially
polarized states,  respectively. 

\section{Halperin state and gluing of charge and spin}

An extension of the Laughlin state to two-component systems in the
plane is
the Halperin state \cite{helv}, given by
\begin{equation}
\Psi_{m m n} (z_{1\uparrow} \cdots z_{N_{\uparrow}},
              z_{1 \downarrow} \cdots z_{N_{\downarrow}})
=\prod_{i<j} (z_{i \uparrow} - z_{j \uparrow})^{m}
\prod_{k<l} (z_{k \downarrow} - z_{l \downarrow})^{m}  \label{eq:halwafu}
\prod_{r<s} (z_{r \uparrow} - z_{s \downarrow})^{n},
\end{equation}
for two components $ \uparrow $ and  $ \downarrow $. It is assumed
that $ N_{\uparrow} = N_{\downarrow} $, {\em i.e.} the state is
unpolarized. It describes fermions if $ m $ is an odd (positive) integer.  (The exponential
factors are omitted for simplicity.) In the case
$ m = n + 1 $,  this state
represents 
a ground spin-singlet state of a system
with spin. If not otherwise specified, we will keep $ m $ and $ n $
general in the following. The pseudo-spin of  two-component systems
will also be called spin for short. In (\ref{eq:halwafu}) we use, as  is
customary,
 the shortened form of the complete wave function, which  would include
spin
vectors and overall antisymmetrization.
The quasihole excitations carry  spin $ 1/[2 (m - n)] $  which
represents the net spin localized in the region around the quasihole. 
They carry $ - 1/(m + n) $
(quasiholes) or $ + 1/(m + n) $ (quasielectrons)
unit of the electric charge.  For
$ S_{z} = 1/[2 (m - n)]  $, where $ S_{z} $ is the $z$ component of the
quasihole spin, and in the case when $m - n = 1$, a single  quasihole excitation can be given by \cite{fdm},
\begin{equation}
\prod_{i=1}^{N_{\downarrow}} (z_{i \downarrow} - w) \Psi_{m m n}, \label{oneqh}
\end{equation}
and analogously for $ S_{z} = - 1/[2 (m - n)] = -1/2 $. For general
$m$ and $n$, (\ref{oneqh}) must be generalized to the one that
includes $|m - n|$ excitations.  
 
The description of the system with each particle having a definite spin 
polarization can be  given  by an effective Chern-Simons theory
\cite{zhanghk} also,
in the so-
called $ U(1) \times U(1) $ formulation with two abelian gauge fields
$ a_{\uparrow}^{\mu} $ and $ a_{\downarrow}^{\mu} $ \cite{eziw}. 
The Chern-Simons
constraints that it contains are
\begin{equation}
\vec{\nabla} \times \vec{a}_{\uparrow} = 
2 \pi (m \rho_{\uparrow} + n \rho_{\downarrow}) \label{eq:halj}
\end{equation}
and
\begin{equation}
\vec{\nabla} \times \vec{a}_{\downarrow} = 
2 \pi (m \rho_{\downarrow} + n \rho_{\uparrow}), \label{eq:hald}
\end{equation}
which are
simple generalizations of the constraint in the Laughlin case
$ (\vec{\nabla} \times \vec{a} = 2 \pi m \rho) $. ( $ \rho_{\uparrow} $ and
$ \rho_{\downarrow} $ are the densities of the up and down electrons, respectively.)
The field theory can be also expressed in terms of 
\begin{equation}
a_{c}^{\mu} = \frac{a_{\uparrow}^{\mu} + a_{\downarrow}^{\mu}}{2}
\; \; {\rm and } \; \;
a_{s}^{\mu} = \frac{a_{\uparrow}^{\mu} - a_{\downarrow}^{\mu}}{2}, \label{defcs}
\end{equation}
which are charge and spin gauge fields, respectively. Their flux is connected with
the charge and spin of the system as can be seen from the previous equations.
The lagrangian density of the theory is then given by
\begin{eqnarray}
\cal L  = & &     \sum_{\sigma} 
i \Psi_{\sigma}^{\dagger}(\partial_{0} + i (a_{c 0}+ \sigma a_{s 0}) 
- i A_{0}) \Psi_{\sigma} +
     \sum_{\sigma i} 
      \frac{1}{2M}
  \Psi_{\sigma}^{\dagger} (\partial_{i}+ i (a_{c i}+ \sigma a_{s i})- i
  A_{i})^{2} \Psi_{\sigma}  \nonumber \\
    & & + \frac{1}{4 \pi} \frac{2}{m+n} \epsilon^{\mu \nu \label{eq:halden} 
\lambda} a_{c \mu} \partial_{\nu} a_{c \lambda} + 
\frac{1}{4 \pi} \frac{2}{m-n} \epsilon^{\mu \nu 
\lambda} a_{s \mu} \partial_{\nu} a_{s \lambda},
\end{eqnarray}
where $ i =x, y $ and $ \Psi_{\sigma} $ $ (\sigma = + \; {\rm and} \; - \; {\rm for} \uparrow {\rm and} \downarrow) $ is a bosonic field that represents the electron
field up to the statistical transformation embodied in $ {\cal L} $.
(For simplicity we omitted the interaction and Zeeman-energy term.)

A vortex excitation of finite energy, in a charged bosonic system,
can occur only if it is accompanied by an increase in the flux of the
gauge fields ($ a_{\downarrow}^{\mu} $ or $ a_{\uparrow}^{\mu} $) in
the amount 
of an integer number of flux quanta; the flux quantum is ${2 \pi}$ in our units.
To describe the excitations, we can integrate Eqs. (\ref{eq:hald}) and
 (\ref{eq:halj}), and relate
the total changes in the flux of $ a_{c} $ and $ a_{s} $ fields, $ \Phi_{c} $ and
$ \Phi_{s} $, respectively, to the total local change in the charge and spin,
 $ Q_{c} $ and $ S_{z} = Q_{s}/2 $, respectively:
 \begin{equation}
 - Q_{c} + \frac{1}{2 \pi} \frac{2}{m + n} \Phi_{c} = 0\; {\rm and} \;
 - Q_{s} + \frac{1}{2 \pi} \frac{2}{m - n} \Phi_{s} = 0.
 \end{equation}
 Taking four choices for $ ( \Phi_{c}, \Phi_{s} ) $ : $ (\pi, \pi),
 (\pi, -\pi),(-\pi, \pi),$ and $(-\pi, -\pi) $ gives four elementary 
 excitations with the same quantum numbers as in the wave function
approach; $ (Q_{c}, Q_{s}) = (1/(m+n), 1/(m-n)), (1/(m+n), - 1/(m-n)),
(- 1/(m+n), 1/(m-n)), (- 1/(m+n), - 1/(m-n)) $, respectively.
  Note that the half-flux quantum changes in
  $ \Phi_{c} $ or $ \Phi_{s} $ cannot occur independently, which is 
  an expression of the charge-spin confinement, {\em i.e.} ``gluing'', in
  the bulk of the system. Large vortices carry fluxes $ (a \pi, b \pi)
$ where $ a + b $ is even.  The Aharonov-Bohm-Berry phase $\theta$ for
an interchange of two vortices, with charges and fluxes,
$(Q_{c}^{1}, Q_{s}^{1})$ and $(\Phi_{c}^{1}, \Phi_{s}^{1})$ for 
vortex 1, and $(Q_{c}^{2}, Q_{s}^{2})$ and $(\Phi_{c}^{2}, \Phi_{s}^{2})$ for  vortex 2,  is given by
\begin{equation}
\frac{\theta}{\pi} = Q_{c}^{1} \frac{\Phi_{c}^{2}}{2 \pi} +  \label{staang}
                     Q_{s}^{1} \frac{\Phi_{s}^{2}}{2 \pi}.
\end{equation}
 If we interchange two elementary vortices of the same
kind the phase or statistical angle is
\begin{equation}
\theta = \frac{m}{m^{2} - n^{2}} \pi. \label{staangsame}
\end{equation}
We may  conclude also that the excitations with fluxes
$ (\pm \pi (m + n), \pm \pi (m - n)) $ are equivalent to one-electron
excitations, which can be particles or holes of either spin, $ S_{z} =
\pm 1/2 $.

Even for  systems as simple as these Halperin states, we may find the lattice structure of excitations
mentioned in the introduction. The excitations can be represented as
 vectors associated with points of a two-dimensional lattice
with
components being integers, $ a = \Phi_{c}/\pi $ and $ b = \Phi_{s}/\pi $, in a basis. The
expression for the statistical angle (\ref{staang}) defines the scalar
product
in this lattice. For two vectors $ {\bf v}_{1}=(a_{1}, b_{1}) $ and 
$ {\bf v}_{2}=(a_{2}, b_{2}) $, it is
\begin{equation}
{\bf v}_{1} \cdot {\bf v}_{2}  = \frac{a_{1} a_{2}}{2 (m + n)} + 
\frac{b_{1} b_{2}}{2 (m - n)}. \label{scproduct}
\end{equation}
We see that the basis used here, $ {\bf e}_{1} = (1, 0)$, $ {\bf
e}_{2} = (0, 1)$, is
orthogonal;  however, these vectors, ${\bf e}_{1}, {\bf e}_{2}$, are not
in the lattice as they fail to satisfy the condition $a + b =$ even.
The condition expresses the fact that the excitation lattice is a
special composite of two one-dimensional lattices, one for charge and
one for spin. (We defer a complete description of the gluing
construction to Sec. III {\bf A}, and, here, we give only a simplified
version of it.)
The charge and spin lattice points are $ (a,0), a = 0, \pm 1, \pm 2, \ldots $ and $ (0,b), b = 0, \pm
1, \pm 2, \ldots $, respectively and, in general, they do not belong to
the excitation lattice. The gluing of the two lattices is specified by
a  rule that we impose on possible combinations of points from the
lattices. In our case the rule is $a + b =$ even.

One way to define the charge and spin lattices is to consider the
sublattice of the excitation lattice connected with the order-parameters of
the system \cite{read}. For Halperin states the order parameter
excitations are specified by vectors: $ {\bf v} = ((m + n) i, (m - n)
k) $ for which $i + k = $ even. The excitations represent multiples of
one-electron excitations which can be particle or hole with $ S_{z} =
\pm 1/2 $. The sublattice that they make can be defined as  dual
to the excitation lattice, {\em i.e.} the one whose vectors have
integer scalar products (see Eq. (\ref{scproduct})) with all other
vectors of the excitation lattice. Then, it is appropriate to define
the charge and spin lattices in this sublattice as those with charge only
and spin only order-parameter excitations, {\em i.e.} those defined by
vectors:
${\bf v}_{c} = ((m + n)i, 0),\; i =$ even, and ${\bf v}_{s} = (0, (m -
n) k), \; k =$ even, respectively. Their dual lattices are the ones
that we used in the preceding description of the gluing.

The Halperin states are the simplest example of the gluing construction.
(A desription of the lattices specialized to the (3,3,1) Halperin
state was also given in Ref. \cite{mimi}.)

  By some standard transformations, see, for example, \cite{zhang},
   we can transform $ {\cal L} $ into the
  one that describes the dual Chern-Simons theory. In the dual theory in the
  Laughlin case, the vortex excitations ({\em i.e.} fluxes previously) are
now to be
  viewed as particles, and what was the particle current density  becomes flux of
  some gauge field. In its low-energy limit the lagrangian of the dual theory
  in our case is
\begin{eqnarray}
  {\cal L}  = & &- {\cal J}_{\mu}^{s} A_{s}^{\mu} - \frac{(m - n)}{2 \pi}
 \epsilon^{\mu \nu \lambda} A_{s \mu} \partial_{\nu} A_{s \lambda}  \nonumber \\
 &  & - {\cal J}_{\mu}^{c} A_{c}^{\mu} - \frac{(m + n)}{2 \pi}
\epsilon^{\mu \nu \lambda} A_{c \mu} \partial_{\nu} A_{c \lambda}   \nonumber \\    
&  & + \frac{1}{\pi} \epsilon^{\mu \nu \lambda} A_{\mu}^{ext} \partial_{\nu} 
A_{c \lambda}.
\end{eqnarray}
The original charge and spin current densities
, $ J_{\mu}^{c} $ and  $ J_{\mu}^{s}$, respectively,  
are now given by
\begin{equation}
J_{c}^{\mu} = \epsilon^{\mu \nu \lambda} 
\frac{\partial_{\nu} A_{c \lambda}}{\pi} \; {\rm and} \;
J_{s}^{\mu} = \epsilon^{\mu \nu \lambda} 
\frac{\partial_{\nu} A_{s \lambda}}{\pi}.
\end{equation}
The currents, $  {\cal J}_{\mu}^{c} $ and $ {\cal J}_{\mu}^{s} $, represent
the charge and spin current densities of the quasiparticles, respectively. 
These currents are measured in the units of charge and spin equal to the
elementary quasiparticle charge and spin. To have
a complete low-energy theory of the FQHE we must also impose the gluing among
these quasiparticles, {\em i.e.} specify which fused combinations of them are
allowed.

In the SU(2)-invariant case, where $ m = n + 1$,
this dual theory can be replaced by  an explicitly SU(2)-invariant
field theory  by
introducing, instead of abelian gauge field $ A_{s \mu} $,  a
nonabelian  gauge field
$ {\cal A}_{\mu} = {\cal A}_{\mu}^{a} \tau^{a} $ where $ \tau^{a}, a = 1,2,3 $
are Pauli matrices. Then, instead of the Chern-Simons term 
\begin{equation}
\frac{(m - n)}{2 \pi} \epsilon^{\mu \nu \lambda} A_{s \mu} \label{cscs}
\partial_{\nu} A_{s \lambda}
\end{equation}
with
$ A_{s \mu} $, in the SU(2)-gauge-invariant theory, we have $ {\rm SU(2)}_{k}$
Chern-Simons term where $ k = 1$ \cite{sam,witten}, {\em i.e.}
\begin{equation}
\frac{k}{4 \pi} \epsilon^{\mu \nu \lambda}      \label{subs}
tr ( {\cal A}_{\mu} \partial_{\nu} {\cal A}_{\lambda} +
\frac{2}{3} {\cal A}_{\mu} {\cal A}_{\nu} {\cal A}_{\lambda}).
\end{equation}
This identification of two theories is possible because
${\rm SU(2)}_{k=1}$ Chern-Simons theory has only excitations with abelian statistics
\cite{bafra} and therefore, can be formulated also in the abelian way with one
abelian gauge field.

\section{Spin-singlet hierarchy of Halperin states and gluing theory}

\subsection{Hierarchy of Halperin states and gluing theory}

It is straightforward to derive the dual Chern-Simons theory lagrangian of the 
hierarchy of the Halperin states where at each level quasiparticles combine into
a new Halperin state. It is a simple generalization of the lagrangian
for the one-component hierarchy given in Ref. \cite{blwe}. The part of the lagrangian for the first-level
hierarchy that we will immediately use represents constraint
conditions on (uniform) charge current densities which define the
ground state.  For the sake of clarity we set current densities of
vortex excitations to zero. The expression for this  part of the
lagrangian is
\begin{eqnarray}
{\cal L =}     \label{duall}
&  &+ \epsilon_{\mu \nu \lambda} \frac{\large 1}{\large 2 \pi} 
A^{\mu}_{ext} \partial^{\nu} (A_{0 \uparrow}^{\lambda} + A_{0 \downarrow}^{\lambda})  \nonumber \\
&  & - \frac{m_{0}}{4 \pi} \epsilon_{\mu \nu \lambda} 
A_{0 \uparrow}^{\mu} \partial^{\nu} A_{0 \uparrow}^{\lambda}
- \frac{m_{0}}{4 \pi} \epsilon_{\mu \nu \lambda} 
A_{0 \downarrow}^{\mu} \partial^{\nu} A_{0 \downarrow}^{\lambda}
- \frac{n_{0}}{2 \pi} \epsilon_{\mu \nu \lambda} 
A_{0 \uparrow}^{\mu} \partial^{\nu} A_{0 \downarrow}^{\lambda}
- \frac{n_{0}}{2 \pi} \epsilon_{\mu \nu \lambda} 
A_{0 \downarrow}^{\mu} \partial^{\nu} A_{0 \uparrow}^{\lambda}  \nonumber \\
&  & - \frac{1}{2 \pi} \epsilon_{\mu \nu \lambda}
A_{0 \uparrow}^{\mu} \partial^{\nu} A_{1 \downarrow}^{\lambda} 
- \frac{1}{2 \pi} \epsilon_{\mu \nu \lambda}
A_{0 \downarrow}^{\mu} \partial^{\nu} A_{1 \uparrow}^{\lambda}  \nonumber \\
&  & - \frac{m_{1}}{4 \pi} \epsilon_{\mu \nu \lambda} 
A_{1 \uparrow}^{\mu} \partial^{\nu} A_{1 \uparrow}^{\lambda}
- \frac{m_{1}}{4 \pi} \epsilon_{\mu \nu \lambda} 
A_{1 \downarrow}^{\mu} \partial^{\nu} A_{1 \downarrow}^{\lambda}
- \frac{n_{1}}{2 \pi} \epsilon_{\mu \nu \lambda} 
A_{1 \uparrow}^{\mu} \partial^{\nu} A_{1 \downarrow}^{\lambda}
- \frac{n_{1}}{2 \pi} \epsilon_{\mu \nu \lambda} 
A_{1 \downarrow}^{\mu} \partial^{\nu} A_{1 \uparrow}^{\lambda},  
\end{eqnarray}
where $ A^{\mu}_{ext} $ represents an external electromagnetic field,
numbers 0 and 1 denote the levels of the hierarchy, and the previous $m$ and
$n$ are now $m_{0}$ and $n_{0}$. The equations of motion obtained from
(\ref{duall}) are
\begin{eqnarray}
& N_{\phi} = m_{0} N_{\sigma}^{0} + 
n_{0} N_{-\sigma}^{0} + \alpha  N_{-\sigma}^{1}, & \nonumber \\ \label{algeq}
& 0 = N_{\sigma}^{0} + m_{1} \alpha N_{\sigma}^{1} + n_{1} \alpha  N_{-\sigma}^{1} &
\end{eqnarray}
with $ \sigma = \uparrow {\rm or} \downarrow $, $ N^{0}_{\sigma} $ and
$ N^{1}_{\sigma}$ denote the total numbers of electrons and
quasiparticles, respectively, $ \alpha = +1 $ for quasiholes or $
\alpha = -1$ for quasielectrons, and $ N_{\phi}$ is the number of flux
quanta through the system. $ m_{1}$ is an even integer because we
consider the quasiparticles as bosons. 
 
 By defining new gauge fields, analogously 
 to what we did for a single Halperin state (\ref{defcs}),  the lagrangian
 density (\ref{duall}) can be rewritten in a form with charge and spin variables
 only,
 \begin{eqnarray}
\cal L =  &  &-\frac{(m_{0}-n_{0})}{2 \pi}
\epsilon_{\mu \nu \lambda} A^{\mu}_{s 0} \partial^{\nu}
A^{\lambda}_{s 0}
-\frac{(m_{0}+n_{0})}{2 \pi}
\epsilon_{\mu \nu \lambda} A^{\mu}_{c 0} \partial^{\nu}
A^{\lambda}_{c 0}
+\epsilon_{\mu \nu \lambda} \frac{A^{\mu}_{ext} \partial^{\nu}
A^{\lambda}_{c 0}}{\pi} \nonumber \\
&  & -\frac{(m_{1}-n_{1})}{2 \pi}
\epsilon_{\mu \nu \lambda} A^{\mu}_{s 1} \partial^{\nu}
A^{\lambda}_{s 1}
-\frac{(m_{1}+n_{1})}{2 \pi}
\epsilon_{\mu \nu \lambda} A^{\mu}_{c 1} \partial^{\nu} A^{\lambda}_{c 1}
 \nonumber \\
&  & - \epsilon_{\mu \nu \lambda} 
\frac{a^{\mu}_{c 0} \partial^{\nu} A^{\lambda}_{c 1}}{\pi}  \label{lef}
- \epsilon_{\mu \nu \lambda} 
\frac{A^{\mu}_{s 0} \partial^{\nu} A^{\lambda}_{s 1}}{\pi}
 - {\cal J}^{c 0}_{\mu} A_{c 0}^{\mu} - {\cal J}^{s 0}_{\mu} A_{s 0}^{\mu} 
 - {\cal J}^{c 1}_{\mu} A_{c 1}^{\mu} - {\cal J}^{s 1}_{\mu} A_{s 1}^{\mu}, 
\end{eqnarray}
where we  included also the vortex excitations with respect to 0 and 1 level
described by current densities $ {\cal J}^{c 0}_{\mu} $ and 
$ {\cal J}^{s 0}_{\mu} $, and $ {\cal J}^{c 1}_{\mu} $ and
$ {\cal J}^{s 1}_{\mu} $, respectively. As before the current densities
are measured in the units of the corresponding quasiparticle charge
and spin. The form of the lagrangian in (\ref{lef}) suggests the definition 
of two matrices,
 charge 
\begin{equation} 
 \Lambda_{c}=\left( \begin{array}{cc}
              m_{0}+n_{0}  &  1          \\
                1     &  m_{1}+n_{1}             \label{chargem}
            \end{array}   \right),
\end{equation}
and spin
\begin{equation}
 \Lambda_{s}=\left( \begin{array}{cc}
              m_{0}-n_{0}  &  1          \\
                1     &  m_{1}-n_{1}
            \end{array}   \right).                \label{spinm}
\end{equation}
They are analogous to the one-dimensional ones, $ m_{0} + n_{0} $ and
$ m_{0} - n_{0} $, for the Halperin state in Sec. II.

The elementary excitations from each level of the constructed hierarchy are of
the type discussed in the section on the Halperin state. Therefore, if we introduce
vectors
\begin{equation}
 N^{c} = \left( \begin{array}{c}
                      N_{0}^{c} \\ N_{1}^{c} 
                     \end{array} \right) 
                      \; \;
            {\rm and}           \label{intvec}
                      \; \;
 N^{s} =   \left( \begin{array}{c}
                      N_{0}^{s} \\ N_{1}^{s}
                     \end{array} \right) 
\end{equation}
with entries denoting how many, with respect to the level 0 or 1, quasiparticles
that carry only charge $ (c) $ or only spin $ (s) $ are created, not any pair
of these vectors represents a system excitation. (The entries can be negative denoting the
number of quasiparticles with opposite charge or spin.)  Because of the
gluing of these particles in the bulk of the system, only integral linear 
combinations of the following elementary excitations from the zeroth level of
the hierarchy:
\begin{equation}
\left\{ \left( \begin{array}{c}
                      1 \\ 0 
                     \end{array} \right)_{c},
                      \left( \begin{array}{c}
                      1 \\ 0
                      \end{array} \right)_{s} \right\} \; {\rm and} \;
                      \left\{ \left( \begin{array}{c}
                      1 \\ 0                   \label{zerothv}
                      \end{array} \right)_{c},
                       \left( \begin{array}{c}
                      -1 \\ 0
                      \end{array} \right)_{s} \right\},
\end{equation}
and
\begin{equation}
\left\{ \left( \begin{array}{c}
                      0 \\ 1
                     \end{array} \right)_{c},
                      \left( \begin{array}{c}
                      0 \\ 1
                      \end{array} \right)_{s} \right\} \; {\rm and} \;
                      \left\{ \left( \begin{array}{c}  \label{firstv}
                      0 \\ 1
                      \end{array} \right)_{c},
                       \left( \begin{array}{c}
                      0 \\ -1
                      \end{array} \right)_{s} \right\}
\end{equation}
from the first level of the hierarchy, are allowed. If we denote by $ x $
 a vector
from the charge part and by $ y $ a vector from the spin part, the operations
allowed on the vectors in (\ref{zerothv}) and (\ref{firstv}) are
\begin{equation}
1. \; \; \{x_{1}, y_{1}\} + \{x_{2}, y_{2}\} \stackrel{\rm def}{=}
 \{ x_{1} + x_{2}, y_{1} + y_{2}\}
\end{equation}
and if $ \alpha $ is an integer
\begin{equation}
2. \; \; \alpha \{x, y\} \stackrel{\rm def}{=} \{\alpha x, \alpha y\}.
\end{equation}
These are familiar operations for a direct sum of two vector spaces.
In addition, if the scalar product is defined by
\begin{equation}
\{x_{1},y_{1}\} \cdot \{x_{2},y_{2}\} = x_{1} \cdot x_{2}
+ y_{1} \cdot y_{2}
\end{equation}
the direct sum is orthogonal.

To any excitation we can assign the corresponding change in fluxes in the dual
Chern-Simons theory (\ref{lef}). Then the Aharonov-Bohm-Berry phase for an 
interchange of two identical excitations gives the statistical angle for
that excitation and it is equal to
\begin {equation}
\frac{\theta}{\pi} = N^{c} \frac{\Lambda_{c}^{-1} N^{c}}{2}
+N^{s} \frac{\Lambda_{c}^{-1} N^{s}}{2}.    \label{staangsecond}
\end{equation}
The charge density in the dual Chern-Simons theory (\ref{lef}) can be read off
as a coefficient in front of $ A_{\mu}^{ext} $ and is equal to
\begin{equation}
\rho = \frac{1}{\pi} \epsilon_{\mu \nu \lambda} \partial^{\nu} A_{0 c}^{\lambda}.
\end{equation}
(As before in the theory a unit of charge corresponds to a half of the flux
quantum.) Therefore the charge of the excitation is given by
\begin{equation}
Q_{c} = (\Lambda^{-1}_{c})_{0 I} N_{I}^{c}, \label{eqcharge}
\end{equation}
and analogously the spin by
\begin{equation}
S_{z} = (\Lambda^{-1}_{s})_{0 I} N_{I}^{s} \frac{1}{2}. \label{eqspin}
\end{equation}
We can rewrite the coupling between the charge density and external field 
\cite{wenzee} as
\begin{equation}
\rho A^{ext} = \frac{1}{\pi} \epsilon_{\mu \nu \lambda} A^{\mu}_{ext}
\partial^{\nu} A_{0 c}^{\lambda} = \frac{1}{\pi} \epsilon_{\mu \nu \lambda}
 A^{\mu}_{ext} \sum_{i=0,1} V_{i} \partial^{\nu} A_{i c}^{\lambda}, 
\end{equation}
where
\begin{equation}
  V =  \left( \begin{array}{c}
                      1 \\  0
                      \end{array} \right). 
\end{equation}
Along with the $ S $ transformations the vector $ V $ will change (to leave the coupling 
invariant and characterize the nature of the coupling between the external field
$ A_{\mu}^{ext} $ and internal gauge fields in a new basis). Therefore, we may
rewrite the physical quantity $ Q_{c} $ in a basis-independent way as
\begin{equation}
Q_{c} = V \Lambda_{c}^{-1} N^{c},
\end{equation}
and similar arguments lead to
\begin{equation}
S_{z} = V \Lambda_{s}^{-1} N^{s} \frac{1}{2}.
\end{equation}

The given description of the excitations may lead us to the conclusion that
they are best described by an orthogonal direct sum of two two-dimensional lattices, one
for spin $ L_{s}^{*} $ and one for charge $ L_{c}^{*} $. With fixed bases
 in them,
$ \{ e_{\alpha}^{c}; \alpha = 0,1 \} $ and
$ \{ e_{\beta}^{s}; \beta = 0,1 \} $,  
 the excitations are given by the integer vectors
(\ref{intvec}), 
 and their scalar products
 are defined on these lattices with two Gram matrices,
$ (\Lambda^{-1}_{c}/2)_{\alpha \beta} = 
e_{\alpha}^{c} \cdot e_{\beta}^{c} $ and
$ (\Lambda^{-1}_{s}/2)_{\alpha \beta} = 
e_{\alpha}^{s} \cdot e_{\beta}^{s} $. 
The Gram matrices are read off from the expression
(\ref{staangsecond})  for the statistical angle, which plays the role
of the scalar product for the vectors (\ref{intvec}). But, as in the case of the
systems with the Halperin ground states, the excitation lattice is the
result of the gluing construction, and it is a sublattice of this
direct sum. Because of the requirement for the flux quantization, only
vectors (\ref{intvec}) whose components satisfy conditions,
\begin{equation}
N_{0}^{c} \pm N_{0}^{s} =  {\rm even\; \; and} \; \;  N_{1}^{c} \pm N_{1}^{s} =
{\rm even}, \label{conds}
\end{equation}
 belong to the excitation lattice. 

The order-parameter lattice is a
sublattice of the excitation lattice, and can be defined in the
following way. (Note that we use the words the order-parameter lattice
instead of the condensate lattice of Ref. \cite{read}.)
If, in the standard-hierarchy basis, we define  the vector
$W$,
\begin{equation}
  W =  \left( \begin{array}{c}
                      0 \\  1
                      \end{array} \right), 
\end{equation}
the numbers
\begin{equation}
Q^{qp} = W \Lambda_{c}^{-1} N^{c} \; \; {\rm and} \; \;
S_{z}^{qp} =(W \Lambda_{s}^{-1} N^{s})/2,
\end{equation}
have the meaning of the quasiparticle charge and spin. Then the 
order-parameter lattice consists of points whose vectors $N^{c}$ and $N^{s}$
satisfy conditions $Q \pm 2 S_{z} =$
 even and $Q^{qp} \pm 2 S_{z}^{qp}  =$ even (where $Q, 2 S_{z},
Q^{pq}, {\rm and} \; 2 S_{z}^{qp}$ are integers). Using this definition
and the definition of the scalar product (\ref{staangsecond}), we can see
that the dual of the order-parameter lattice is the excitation
lattice. Namely, all vectors $N^{c}$ and $N^{s}$, which have integer
scalar products with all vectors of the order-parameter lattice,
satisfy the conditions (\ref{conds}). 

Two sublattices can be defined in the order-parameter lattice:  the
charge lattice  $L_{c}$, for which $\; Q = {\rm even}, \; 2 S_{z} = 0, \; Q^{qp} =
{\rm even, \; and} \; 2 S_{z} = 0$, and  the spin lattice  $L_{s}$, for which
$\; Q = 0, \; 2 S_{z} = {\rm even}, \; Q^{qp} = 0, \; {\rm and} \; 2 S_{z} = {\rm
even}.$ The dual charge lattice $ L_{c}^{*}$ consists of arbitrary
integer-component $N^{c}$ vectors, and $N^{s}  =  \left( \begin{array}{c}
                      0 \\  0
                      \end{array} \right)$, and the dual spin lattice
$L_{s}^{*}$ has $N^{c} =  \left( \begin{array}{c}
                      0 \\ 0 
                      \end{array} \right)$
and arbitrary integer-component $N^{s}$ vectors. We used these two lattices, $L_{c}^{*}$ and
$L_{s}^{*}$, to describe the gluing construction. Because their Gram
matrices are $\Lambda_{c}^{-1}/2$ and
$\Lambda_{s}^{-1}/2$, the Gram matrix for the  charge lattice $
L_{c}$ is $2 \Lambda_{c}$, and for the  spin lattice $ L_{s}$ is $2
\Lambda_{s}$ \cite{glue}. 

The gluing construction of lattices can be defined in a
basis-independent  way on a small number of vectors. In the following we will
define
\cite{glue} such a construction for an integral lattice {\em i.e.} the
one with an integer scalar product or integer Gram matrix \cite{glue}. 
The lattice  $ L $ constructed by the gluing  theory contains a sublattice 
which is a direct sum 
\begin{equation}
L_{1} \oplus  L_{2} 
\end{equation}
of two  integral sublattices $ L_{1} \; \;{\rm and} \; \;  L_{2} $. Any vector of $ L $ can be
written as
\begin{equation}
  {\bf y} = {\bf y}_{1} \oplus  {\bf y}_{2},   \label{decomp}
\end{equation}
where each component ${\bf y}_{i} $ belongs to $ L_{i}^{*} $, 
 the dual lattice of $ L_{i} $. 
Therefore, $ {\bf y}_{i} $ is not necessarily in   
$ L_{i} $. To classify candidates for $ {\bf y}_{i} $ we may consider
the sets obtained
by adding to each $ {\bf y}_{i} $ all vectors from $ L_{i} $. Then $
{\bf y}_{i} $ in
 (\ref{decomp}) are representatives of the cosets of $ L_{i} $ in $
L_{i}^{*} $. $ {\bf y}$'s must have integer scalar products with one another, and are
closed under addition modulo $L_{1} + L_{2}$. They
 are known under the name glue vectors.
 
We may identify our order-parameter lattice construction,
with the basis independent, abstract notion of the gluing of two
integral lattices $ L_{c}$ and $L_{s}$ (which play roles of $L_{1}$
and $L_{2}$ in the preceding paragraph).
As we already said, in the standard-hierarchy basis the matrices 
$ 2 \Lambda_{c} $ and $ 2 \Lambda_{s} $ (see Eq. (\ref{chargem}) and
(\ref{spinm})) are the Gram matrices of the two integral lattices $
L_{c} $ and $ L_{s}$. If $ {\bf a}_{1c}$ and $ {\bf a}_{2c}$ denote
the
standard-hierarchy basis vectors of the lattice $ L_{c}$, the matrix
$ 2 \Lambda_{c}$ encodes  information about their scalar products
in the following way \cite{glue}:
\begin{equation}
2 \Lambda_{c} = \left( \begin{array}{cc}
        {\bf a}_{1c} \cdot {\bf a}_{1c} & {\bf a}_{1c} \cdot {\bf
a}_{2c} \\
        {\bf a}_{1c} \cdot {\bf a}_{2c} & {\bf a}_{2c} \cdot {\bf
a}_{2c}
        \end{array} \right).
\end{equation}
Therefore, if $ {\bf e}_{1} = (1, 0)$ and $ {\bf e}_{2} = (0, 1)$ are
the orthogonal basis vectors with unit norms, the vectors
$ {\bf a}_{1c} $ and $ {\bf a}_{2c} $ can be expressed as
\begin{equation}
{\bf a}_{1c} = \sqrt{2 (m_{0} + n_{0})} {\bf e}_{1},
\end{equation}
and
\begin{equation}
{\bf a}_{2c} = \sqrt{\frac{2}{m_{0} + n_{0}}} {\bf e}_{1} +
               \sqrt{\frac{2 D_{c}}{m_{0} + n_{0}}} {\bf e}_{2},
\end{equation}
where $ D_{c} = \det \Lambda_{c}$.
The dual (excitation)-lattice basis vectors are
\begin{equation}
{\bf b}_{1c} = \sqrt{\frac{1}{2 (m_{0} + n_{0})}} {\bf e}_{1} - 
\sqrt{\frac{1}{2 D_{c} (m_{0} + n_{0})}} {\bf e}_{2},
\end{equation}
and
\begin{equation}
{\bf b}_{2c} = \sqrt{\frac{(m_{0} + n_{0})}{2 D_{c}}} {\bf e}_{2}.
\end{equation}
(They are obtained from the requirement that any dual-lattice vector
$ {\bf v} = p {\bf e}_{1} + q {\bf e}_{2} $, $ p $ and $ q $, in
general, rational, has an integral scalar product with vectors
 $ {\bf w} = k {\bf a}_{1c} + l {\bf a}_{2c} $, $ k $ and $ l $
integer, of the integral lattice $ L_{c} $.) We may check that the
volume of the elementary integral-lattice ($L_{c}$) cell is
$ 4 D_{c} = \det\{2 \Lambda_{c}\} $ times larger than the volume of the 
elementary dual lattice cell, {\em i.e.}
\begin{equation}
\frac{|{\bf a}_{1c} \times {\bf a}_{2c}|}{|{\bf b}_{1c} \times {\bf b}_{2c}|} = 4 D_{c}.
\end{equation}
This is a one way to see that there are $ \det\{2 \Lambda_{c}\} $, {\em
i.e.} the determinant of the integral-lattice Gram matrix, coset
representatives of $ L_{c}$ in $ L_{c}^{*}$ \cite{glue}. To find a
 complete set of them, we require that
\begin{equation}
 0 \leq {\bf v} \cdot {\bf e}_{1} < {\bf a}_{1} \cdot {\bf e}_{1} 
 \; \; {\rm and} \; \;       
 0 \leq {\bf v} \cdot {\bf e}_{2} < {\bf a}_{2} \cdot {\bf e}_{2},
\label{procond}
\end{equation}
that is we look for all that are in the volume of the integral-lattice
($L_{c}$) elementary cell. To give an example, we specialize to the
case for which $ m_{0} + n_{0} = 1$. Taking the conditions
(\ref{procond}) into account, we get
\begin{equation}
\left( \begin{array}{c}
       0 \\ 0
        \end{array} \right)_{c},
\left( \begin{array}{c}
       0 \\ 1
        \end{array} \right)_{c}, \ldots,
\left( \begin{array}{c}
          0  \\  2 (m_{1} + n_{1}) - 3
         \end{array} \right)_{c},
\end{equation}
and
\begin{equation}
\left( \begin{array}{c}
        1 \\ 1
       \end{array} \right)_{c},
\left( \begin{array}{c}
        1 \\ 2
       \end{array} \right)_{c}, \ldots,
\left( \begin{array}{c}
        1 \\ 2 (m_{1} + n_{1}) - 2
       \end{array} \right)_{c}
\end{equation}
as a set of the coset representatives of $ L_{c} $ in $ L_{c}^{*}$.
(There are $ 4 [(m_{1} + n_{1}) - 1] = 4 D_{c}$ of them.) The
components
of the vectors are defined relative to the dual-lattice ($ L_{c}^{*}$)
basis
vectors (${\bf b}_{1c}$ and ${\bf b}_{2c}$).
The analysis for the spin lattices can be repeated in a similar way
and the results only differ from the charge case in that, that we have
to
change the sign of $ n_{i}$, $ i = 0,1 $. For this
special case $ (m_{0} \pm n_{0} = 1) $ the glue vectors from the
definition of the gluing construction are
\begin{eqnarray}
& \left( \begin{array}{c}
      0 \\ 0
       \end{array} \right)_{c} \oplus
 \left( \begin{array}{c}
      0 \\ 0
       \end{array} \right)_{s},
\left( \begin{array}{c}
      1 \\ 1
       \end{array} \right)_{c} \oplus
 \left( \begin{array}{c}
      1 \\ 1
       \end{array} \right)_{s},&  \nonumber \\
& \left( \begin{array}{c}
      0 \\ (m_{1} + n_{1}) - 1
       \end{array} \right)_{c} \oplus
 \left( \begin{array}{c}
      0 \\ (m_{1} - n_{1}) - 1
       \end{array} \right)_{s} \; {\rm and}
& \left( \begin{array}{c}
      1 \\ m_{1} + n_{1}
       \end{array} \right)_{c} \oplus
 \left( \begin{array}{c}
      1 \\ m_{1} - n_{1}
        \end{array} \right)_{s},
\end{eqnarray}
where we also assumed that $ m_{1} + n_{1} $ is an even number. The
last assumption is appropriate, as we will see, in the case of the
spin-singlet constructions.
The glue vectors satisfy the gluing rule that we found considering
the
elementary excitations of the first-level hierarchy; we glue only
those
coset-representative vectors for which components for each hierarchy
level
(in the standard-hierarchy basis) are both even or both odd integers.
They have integral norms, integral scalar products with one another, and are closed
under
addition modulo $ L_{c} \oplus L_{s} $. All other vectors of the
order-parameter lattice $ L $ are obtained by adding the ones from
$ L_{c} \oplus L_{s}$ to these glue vectors. From the order-parameter
 lattice the excitation lattice can be constructed, being dual of the
order-parameter lattice.

\subsection{Spin-singlet condition and first-level spin-singlet hierarchy}

The spin-singlet states of the general two-component hierarchy should have 
an explicitly SU(2)-invariant field-theoretical description. So in that case,
it should be possible to cast the low-energy lagrangian density in (\ref{lef}) in 
an explicitly SU(2)-invariant form. The low-energy form in
(\ref{lef}) contains a possibility to make transformations 
\begin{equation}
 \Lambda^{\prime}_{c} = S_{c}^{\top} \Lambda_{c} S_{c}  \; \; {\rm
and} \; \;
 \Lambda^{\prime}_{s} = S_{s}^{\top} \Lambda_{s} S_{s} , \label{trformation}
\end{equation}
\begin{equation}
 {\cal J}^{\prime}_{c} = S_{c} {\cal J}_{c}  \; \; {\rm and} \; \;
 {\cal J}^{\prime}_{s} = S_{s} {\cal J}_{s} ,
\end{equation}
 and simultaneous
inverse transformations on the gauge fields 
\begin{equation}
 S^{-1}_{c} A_{c} =
A^{\prime}_{c} \; \; {\rm  and} \; \;  S^{-1}_{s} A_{s} = A^{\prime}_{s}
\end{equation}
 as found in the 
polarized case in \cite{read,wenzee,hier}.  The
matrices $ S_{c} $ and $ S_{s} $ must be integer matrices, 
$ \mid \det{S_{c}} \mid = 1 $, 
and $ \mid \det{S_{s}} \mid = 1 $, because they must map the excitation
vectors with integer components (\ref{intvec}) into  vectors with integer
components in a one-to-one fashion
 \cite{read,wenzee,hier}. The form of the 
$ {\rm SU(2)}_{k=1} $ Chern-Simons term in (\ref{subs}) suggests  that, only when
 $ \Lambda^{\prime}_{s} $ is in a diagonal form, we would be able to cast the theory in a
 SU(2)-explicitly-invariant form. Therefore, for any spin-singlet construction,
 we must have this possibility of having a  transformation $ S_{s} $ which converts
 $ \Lambda_{s} $ in a diagonal form. Moreover, repeating the
replacement of (\ref{cscs}) with (\ref{subs})
   in this case, we require that the diagonal entries of 
 the transformed $ \Lambda_{s} $ must be 1 or $-1$, {\em i.e.} that we have a
 theory with as many $ {\rm SU(2)}_{k=1} $ Chern-Simons terms as the number of
 the levels in the hierarchy (including the zeroth one). Only with this
 requirement the SU(2)-invariant theory has abelian statistics, which must be the statistics 
 of the excitations in any hierarchy built up from the Halperin states.

The formulated spin-singlet condition implies that the 
spin lattice $ L_{s} $ and the dual spin lattice $
L_{s}^{*} $, in the spin-singlet hierarchy, should be of the
simple square-lattice kind. This is implied because of the existence
of the basis with the diagonal form of $ \Lambda_{s} $ where the absolute
values of the diagonal matrix are the same. Comparing the Gram matrix
$ 2 \Lambda_{s} $ for the spin lattice $ L_{s} $ and the Gram matrix
$ (1/2) \Lambda_{s}^{-1} $ for the dual spin lattice $ L_{s}^{*} $, we
may conclude that $ L_{s} $ is a square lattice with the basis vectors of
twice the length of those describing $ L_{s}^{*} $. 
 
 In the following, the condition formulated in the first paragraph of
this subsection  will be
 applied to the construction of the first-level spin-singlet hierarchy. We will
 find all two-by-two spin matrices of the form (\ref{spinm}) with 
 $ m_{1} $  even,  for which there exist the integer matrices $ S_{s} $ (with
 $ \mid \det S_{s} \mid = 1 $) that transform them into the diagonal form
 with $ -1$ or 1 on the diagonal. The same transformation that is applied to the
 spin part will be applied to the charge part ($ S_{s} = S_{c} = S $), and we
 will identify the new basis as the basis of Jain's construction.
 
 We generate our spin-singlet hierarchy by successively applying the spin-singlet
 condition on each level of the hierarchy. Therefore, the two diagonal matrices
 that we should consider in the spin part,  $ D^{i}_{s}, i = 1,2 $ ( $ D_{s}^{i} = S^{i \top} \Lambda^{i}_{s}
S^{i} $ ),  are
 \begin{equation}
  D_{s}^{1}=\left( \begin{array}{cc}
              1  &  0          \\
                0     &  -1           
            \end{array}   \right)
  \; \; \; {\rm and} \; \; \;
  D_{s}^{2}=\left( \begin{array}{cc}
              1  &  0          \\
                0     &  1             \label{possibilities}
            \end{array}   \right) ; 
 \end{equation}         
the first diagonal entry is the condition $ m_{0} - n_{0} = 1 $ on the zeroth
level of the hierarchy. (We do not consider the $ m_{0} - n_{0} = -1 $ condition
for a reversed magnetic field because the corresponding cases reduce
to the ones given by (\ref{possibilities})).

The first case, because $ \det \Lambda_{s} = \det D_{s} $, corresponds to the
condition $ m_{1} - n_{1} = 0 $. The transformation matrix $ S_{c} =
S_{s} = S $ in (\ref{trformation}) in this case is
\begin{equation}
S^{1}=\left( \begin{array}{cc}
              1  &  1          \\ \label{transmatrix}
                0     &  -1           
            \end{array}   \right),
 \end{equation}
 and the charge part in the new basis is 
 \begin{equation}
 D_{c}^{1}=\left( \begin{array}{cc}
              m_{0} + n_{0}  &  m_{0} + n_{0} - 1         \\ \label{jcharge}
              m_{0} + n_{0} - 1     &  (m_{0} + n_{0}) + (m_{1} + n_{1}) - 2           
            \end{array}   \right).
 \end{equation}
 ($ D_{s}^{1} $ has all $n$'s with minus sign.) 
 
 To understand the physical meaning of the new basis we specialize to
the case
 $ m_{0} + n_{0} = 1 $, that is the case where the zeroth level is the completely filled lowest 
 Landau level of both spins. Then the charge and spin matrices are
 \begin{equation}
  D_{c}^{1}=\left( \begin{array}{cc}
              1  &  0          \\
                0     &  (m_{1} + n_{1}) - 1          
            \end{array}   \right)
  \; \; \; {\rm and} \; \; \;      
  D_{s}^{1}=\left( \begin{array}{cc}
              1  &  0          \\
                0     &  -1           
            \end{array}   \right).  \label{speccase}
 \end{equation}   
 The latter means that there are two independent spin-singlet systems. If, in
 addition, both $ m_{1} $ and  $ n_{1} $ are chosen to be nonpositive even
 numbers, {\em i.e.} $ m_{1} = n_{1} = - 2 n $ where $ n = 1,2,... $,  the
 second system is described by a Halperin state in a reverse magnetic field.
  The filling fraction is $ 2 + 2/[(m_{1} + n_{1}) - 1] = 
  2 - 2/(4 n + 1) $. The form of the description, given by
(\ref{speccase}) of these  Halperin states of holes, suggests that we are
in the basis of Jain's construction.  Indeed, the form of the general hierarchy
  construction, represented by the matrix $ D_{c}^{1} $ in Eq. (\ref{jcharge}),
  tells us that it is the result of the two operations which characterize
   Jain's construction \cite{jain,read,wenzee}; first combining with one Landau level
  (exemplified in (\ref{speccase})) and then attaching flux to electrons to get composite fermions. Explicitly,
  it is
  \begin{equation}
  D_{c}^{1}=\left( \begin{array}{cc}
              1  &  0          \\
                0     &  (m_{1} + n_{1})-1           
            \end{array}   \right)
  \; \;  + \; \; 
  (m_{0} + n_{0} - 1) \left( \begin{array}{cc}
                                1  &  1          \\
                                1     &  1           
                               \end{array}   \right)  
 \end{equation}         
 with the pseudo-identity matrix (with all entries equal to one), which always
 comes with the spin-independent flux-attaching operation. The number of the flux quanta
 attached is $ (m_{0} + n_{0} - 1)/2 $, which is an integer. (Always, to get physical quantities
 that are measured in the unit of one flux quantum, we have to divide the
 defined matrices of the theory by two because they describe a theory where
 the unit for flux is one half of the flux quantum.)
 
 Having identified which values of $m_{1}$ and $n_{1}$  give 
spin-singlets  and recognized that, in the case $ m_{1} = n_{1} = -2n$, in
 the standard-hierarchy picture we have a quasihole system, we are ready to
 construct the corresponding wave function in the fractional-statistics 
 representation \cite{bih} for this case. If we use numbers $p$ and $q$ to describe the
 statistical angles of the quasiholes of the zeroth level as (see Eq.
(\ref{staang}) and (\ref{staangsame}))  
 \begin{equation}
\theta_{i} = \frac{m_{0}}{m_{0}^{2}-n_{0}^{2}} \pi = p \pi
\end{equation}
for the exchange of the same-spin quasiholes, and
\begin{equation}
\theta_{d} = -\frac{n_{0}}{m_{0}^{2}-n_{0}^{2}} \pi = q \pi
\end{equation}
for the exchange of the opposite-spin quasiholes, the wave function is
of the form \cite{bih}
\begin{eqnarray}
\Psi(z_{1 \uparrow},\ldots,z_{N \downarrow}) =& & 
 \exp\{-\frac{1}{4}\sum|z_{i}|^{2}\} 
\prod_{i<j} (z_{i \uparrow} - z_{j \uparrow})^{m_{0}}   
\prod_{i<j} (z_{i \downarrow} - z_{j \downarrow})^{m_{0}}
\prod_{i,j} (z_{i \downarrow} - z_{j \uparrow})^{n_{0}} \nonumber \\
& &\times \int d^{2} w_{1 \uparrow} \cdots \int d^{2} w_{M \downarrow}
\prod_{i,j} (z_{i \uparrow} - w_{j \downarrow})   
\prod_{i,j} (z_{i \downarrow} - w_{j \uparrow}) \nonumber \\
& & \times \prod_{i<j} |w_{i \uparrow} - w_{j \uparrow}|^{2p}  
\prod_{i<j} |w_{i \downarrow} - w_{j \downarrow}|^{2p}          \label{longwf}
\prod_{i,j} |w_{i \uparrow} - w_{j \downarrow}|^{2q} \nonumber \\
&  & \times \prod_{i<j} 
(\overline{w}_{i \uparrow} - \overline{w}_{j \uparrow})^{2 n}   
\prod_{i<j} 
(\overline{w}_{i \downarrow} - \overline{w}_{j \downarrow})^{2 n}
\prod_{i,j} 
(\overline{w}_{i \downarrow} - \overline{w}_{j \uparrow})^{2 n}
\exp\{- \frac{1}{2(m_{0}+n_{0})} \sum |w_{i}|^{2}\} \nonumber \\
\end{eqnarray}
The pseudo-wave function \cite{bih}, which incorporates the effects 
of fractional statistics and describes the physics of the system 
solely in terms of the quasiparticle coordinates, is 
\begin{eqnarray}
\Psi (\overline{w}_{1 \uparrow},\ldots,\overline{w}_{M \downarrow})=
\prod_{i<j} 
(\overline{w}_{i \downarrow} - \overline{w}_{j \downarrow})^{p + 2 n}
\prod_{i<j} 
(\overline{w}_{i \uparrow} - \overline{w}_{j \uparrow})^{p + 2 n}
\prod_{i<j} 
(\overline{w}_{i \downarrow} - \overline{w}_{j \uparrow})^{q + 2 n}.
\end{eqnarray}
The condition \cite{hamer} for spin-singlet
 states for quasiparticles with fractional statistics \cite{bafra} is
 \begin{equation}
(\sum_{j=1}^{M/2} e(i,[j]) \exp\{- i \theta_{s}\} + 1)
\Psi (\overline{w}_{1 \uparrow},\ldots,\overline{w}_{M \downarrow})
= 0,
\end{equation} 
where $ e(i,[j]) $ denotes an exchange between two particles with opposite
spin. Because $ p - q = 1 $,  we can use
a decomposition of the wave function in the form
\begin{eqnarray}
\Psi (\overline{w}_{1 \uparrow},\ldots,\overline{w}_{M \downarrow})= & &
 \prod_{i<j} 
(\overline{w}_{i \downarrow} - \overline{w}_{j \downarrow})^{p + 2 n -1}
\prod_{i<j} 
(\overline{w}_{i \uparrow} - \overline{w}_{j \uparrow})^{p + 2 n - 1}
\prod_{i<j} 
(\overline{w}_{i \downarrow} - \overline{w}_{j \uparrow})^{p + 2 n - 1} \nonumber \\
& & \times \prod_{i<j} 
(\overline{w}_{i \uparrow} - \overline{w}_{j \uparrow})
\prod_{i<j} 
(\overline{w}_{i \downarrow} - \overline{w}_{j \downarrow}),
\end{eqnarray}
where a spin-independent factor multiplies the filled lowest Landau
level of both spins which is a spin-singlet, and see that it obeys the spin-singlet condition. This coincides with our
intuitive expectation that, in the case of the spin-singlet hierarchy 
construction, on the top of the Halperin (zeroth-level) state we have the
spin-singlet state of quasiparticles.

The effects of the statistics in Eq. (\ref{longwf}) are
 in the product with absolute differences, and, as we know from before,
if the
 densities or total numbers of the (quasi)particles that characterize the ground 
 state are concerned, they are irrelevant \cite{hier}. (That can be showed by an
 application of the Laughlin plasma analogy \cite{laugh} to the wave function.) 
 If we consider the quasiparticles as fermionic and, loosely speaking, pull the
 factor $ \prod_{i<j} 
(w_{i \uparrow} - w_{j \uparrow})   
\prod_{i<j} 
(w_{i \downarrow} - w_{j \downarrow}) $ from the product with the absolute
differences in Eq. (\ref{longwf}) to the product (under the integral
signs) involving the differences of $z$'s and $w$'s, and the factor
$ \prod_{i<j} 
(\overline{w}_{i \uparrow} - \overline{w}_{j \uparrow})   
\prod_{i<j} 
(\overline{w}_{i \downarrow} - \overline{w}_{j \downarrow}) $ to the
product with the differences of $\overline{w}$'s to the $2 n^{th}$ power, effectively the state of quasiparticles will be a Halperin
spin-singlet state. So in this case, we  have also a spin-singlet state on the
top of another one. This is to be expected, because the quasiparticles 
are taken to be fermionic in the lowest Landau level just because of their
way of combining together with respect to the internal spin degree of
freedom \cite{fdm}.

In the second case, with $ D_{s}^{2} $ in (\ref{possibilities}), the spin-singlet
requirement is $ m_{1} - n_{1} = 2 $. The matrix $ D_{c}^{2} $ is the same as in
the previous construction. When $ n_{1} = 0 $ {\em i.e.} $ m_{1} = 2$ (and
$ m_{0} $ and $n_{0}$ are of the Halperin states) $ D_{s}^{2} $ and $ D_{c}^{2} $
describe a simple Jain's construction: two Landau levels are completely 
filled with composite fermions obtained by attaching 
$ (m_{0} + n_{0} - 1)/2 $ flux quanta to electrons. In the
standard-hierarchy  picture this is a quasielectron construction. The filling fraction is
\begin{equation}
\nu = \frac{4}{4 [ \frac{\textstyle{m_{0} + n_{0} - 1}}{\textstyle{2}} ] + 1}.
\end{equation}

\subsection{Arbitrary-level spin-singlet hierarchy}

The starting point for the $ n^{th} $-level-hierarchy construction is the spin matrix
$ \Lambda_{s} $ (in the standard-hierarchy basis):
\begin{equation}
\Lambda_{s}^{n} = \left( \begin{array}{cccccc}
 m_{0}-n_{0}  &     1         &    0       & \cdots  &    0    &     0  \\
      1       &  m_{1}-n_{1}  &    1       & \cdots  &    0    &     0  \\
      0       &     1         & m_{2}-n_{2}& \cdots  &    0    &     0  \\
    \vdots    &   \vdots      &   \vdots   & \ddots  &    1    &     0  \\
      0       &     0         &    0       &    1    &m_{n-1}-n_{n-1}&  1 \\
      0       &     0         &    0       &    0    &    1    & m_{n}-n_{n} 
                     \end{array} \right).
\end{equation}
The spin-singlet condition, formulated in the previous subsection and
easily generalized to any level of the hierarchy, demands that
the diagonal entries are such that $ \mid \det \Lambda_{s} \mid = 1 $. We
satisfy this condition by demanding that, at each level $ i $ of the
hierarchy, $ \mid \det \Lambda_{s}^{i} \mid = 1 $. This means that
the differences $ m_{i} - n_{i} , \; i= 0, \ldots, n $, become fixed numbers 
characterizing the constructions. To determine possibilities we define
matrices $ M_{k}, \; k=0,\ldots, n $, of the form 
\begin{equation}
 M_{k} = \left( \begin{array}{ccccc}
 m_{k}-n_{k}  &     1         &    0           & \cdots            &     0 \\
      1       &  m_{k+1}-n_{k+1}  &    1       & \cdots            &     0 \\
      0       &     1         & m_{k+2}-n_{k+2}& \cdots            &     0 \\ 
    \vdots    &   \vdots      &   \vdots       & \ddots            &     1 \\
      0       &     0         &    0           &       1           & m_{n}-n_{n} 
                     \end{array} \right).
\end{equation}
They satisfy the following recursion relation:           
\begin{equation}
  \det M_{k} = (m_{k} - n_{k}) \det M_{k + 1} - \det M_{k + 2},
\end{equation}
where $ \det M_{n + 1} = 1 $ and $ \det M_{n + 2} = 0 $. Repeatedly
requiring at each level $ i $  of the hierarchy that 
$ \mid \det \Lambda_{s}^{i} \mid = 1 $, we have, at the end, this
constraint on the values of $ m_{n} $ and $ n_{n} $:
\begin{equation}
|\det \Lambda_{s}^{n}|=| \det M_{n} \pm \det M_{n+1} |=
|(m_{n}-n_{n}) \pm 1|=1,
\end{equation}
where the sign depends on the way the lower level was constructed.

Because of the nominal Bose statistics of the quasiparticles in this field
theory construction ($ m_{n} $ is even), and as a consequence of the earlier
requirement, $ m_{n} + n_{n} $ is an even number. The filling fraction
in this construction is given by
\begin{equation}
\nu_{n}= 2 (\Lambda_{c}^{-1})_{00} =
2 \frac{1}
{m_{0} + n_{0} 
        - \frac{\textstyle{1}}{\textstyle{m_{1}+n_{1}}-
        _{\ddots_{ -\frac{\textstyle{1}}{\textstyle{m_{n}+n_{n}}}}}}}.
\end{equation}
Because $ m_{0} + n_{0} $ is an odd number and all the rest are even, the
 general form of the filling fraction is {\em an even over an odd} integer.
 
 At each level of the hierarchy we may constrain the signs of $ m_{n} $ and
 $ n_{n} $, demanding that, in the basis of  Jain's construction we begin
 (combining one Landau level) with a Halperin state. This, as we did in the
 case of the first-level quasihole construction, is not necessary. The
Chern-Simons field theory, that we use, enlarges the principles of the
standard-hierarchy construction (\cite{hald,fdm}), and contains also constructions where ground states of the (quasi-)particle systems can be described by
wave functions
which are not of the form of the Halperin states.
Namely, if we take  the
 function of the filled lowest Landau level of both spins and attach
an even
 number $(2 n)$ of flux quanta, in the direction opposite to the external
magnetic field, we get a spin-singlet state, at the filling
 fraction $ \nu^{-1} = (4n-1)/2 $, which differs from the Halperin states. 
These wave functions, projected to the lowest Landau level, are
included in Jain's constructions. As
 a consequence, this enlarged construction, (with no additional constraints on
  $ m_{n} $'s and $ n_{n} $'s), covers all filling fractions of the
  even integer/odd integer  form.
  
  By induction, the matrix $ S_{n} $, at each level $n$, can be found, which
  diagonalizes the matrix $ \Lambda_{s}^{n} $ and gives the corresponding charge
  matrix in the basis of  Jain's construction. We assume the form of
  $ S_{n}^{\top}$, a $ (n+1) \times (n+1) $ matrix, to be
  \begin{equation}
  \label{forma}
S^{\top}_{n} = \left( \begin{array}{cccccc}   \label{eq:transm}
  1       &     0         &    0          &  0           & \cdots  &     0 \\
  1       &    -1         &    0          &  0           & \cdots  &     0 \\
  1       &    -1         &  (-1)^{k_{2}} &  0           & \cdots  &     0 \\
  1       &    -1         &  (-1)^{k_{2}} & (-1)^{k_{3}} & \cdots  &     0 \\
\vdots    &   \vdots      &    \vdots     &   \vdots     & \ddots  &  \vdots \\
  1       &     -1        &   (-1)^{k_{2}}&  (-1)^{k_{3}}& \cdots  &    z 
                     \end{array} \right),
\end{equation}
where each $ k_{i}, \; i=2,\ldots, n-1 $, can be 0 or 1. $ S_{n-1}^{\top} $, a
$ n \times n$ matrix, lies in the upper left corner of $ S_{n}^{\top} $.
 If, for some choice of $ (k_{2},\ldots, k_{n-1}) $, $ S_{n-1}^{\top} $
 diagonalizes given $ \Lambda_{s}^{n-1}$ matrix for the $(n-1)^{th}$ level of
 the hierarchy (so that the diagonal entries are $+1$ and $-1$), and
 $S_{n}^{\top}$ diagonalizes in the same way $ \Lambda_{s}^{n} $, we will
 show that $ \mid z \mid = 1 $ and the sign of $ z $ is uniquely determined by
 the way the lower level is constructed. From the form of $ S_{n}^{\top} $, 
 (\ref{forma}), and the previous assumptions follows that
 \begin{equation}
D_{s}= S_{n}^{\top} \Lambda_{s}^{n} S_{n}=
\left( \begin{array}{cc}
              D_{s}(n \times n) & \begin{array}{c}
                                       0 \\
                                    \vdots \\
                                       0  \\
                                    e + z (-1)^{k_{n-1}}
                                  \end{array}     \\
 \begin{array}{cccc}     
  0 & \cdots & 0 & e + z (-1)^{k_{n-1}}
 \end{array}                     & e + 2 z (-1)^{k_{n-1}} + (m_{n}-n_{n})z^{2}
       \end{array} \right),
\end{equation}
where $ D_{s}(n \times n) $ is a $ n \times n $ diagonal matrix with all
diagonal entries +1 or $-1$, and $e$ represents its last diagonal entry. From 
the constraint that the off-diagonal elements of $ D_{s} $ are also zero follows 
that $ z = (-1)^{k_{n-1}+1} e $ ({\em i.e.} $ \mid z \mid = 1$).

\section{Generalized spin-singlet hierarchy}

We may  consider also constructions  for ground states of spin-singlet
FQHE systems where, at some stage of the hierarchy, elementary quasiparticles
of a Halperin state pair into spinless Laughlin quasiparticles and make
a Laughlin state. This situation, in our formalism, in the case of the first-level spin-singlet hierarchy, is described by the following matrix for the
charge degrees of freedom:
\begin{equation}
 \Lambda_{c}=\left( \begin{array}{cc}
              m_{0} + n_{0}  &  2        \\ \label{ghcharge}
              2   &   4 p          
            \end{array}   \right),
 \end{equation}
 where $ m_{0} $ and $ n_{0} $ are of the Halperin zeroth level state, and
 $ p $ is an arbitrary integer, and matrix 
 \begin{equation}
 \Lambda_{s}=\left( \begin{array}{cc}
              1  &  0        \\ \label{ghspin}
              0   &   0          
            \end{array}   \right)
 \end{equation}
 for the spin degrees of freedom. (The factor 2 in the charge matrix, for the
 part that describes the new construction, which is similar to the 
standard-hierarchy construction in the polarized case, is again a consequence of the
 choice for the unit of flux in our formalism.) The zero on the
diagonal of $ \Lambda_{s}$ is a consequence of the fact that the
description of the localized quasiparticle excitations, in this
system, can be given by a three-dimensional lattice.

%
 
 Again, by applying the matrix transformation $ S $ on both spin and charge
 part simultaneously we may find different expressions of the same construction.
 In the special case when $ S $ is given by (\ref{transmatrix}), the charge and
 spin matrix are
 \begin{equation}
  J_{c}=\left( \begin{array}{cc}
              4n + 1  &  4n - 1          \\
              4n - 1     &  4n - 3 + 4p          
            \end{array}   \right)
  \; \; \; {\rm and} \; \; \;     
  J_{s}=\left( \begin{array}{cc}
              1  &  1          \\
                1    &   1           
            \end{array}   \right), \label{macs} 
 \end{equation}
where $m_{0} + n_{0} = 4n + 1$.
 When $ n = 1 $ ({\em i.e.} in the previous hierarchy construction we begin with
 the $ \nu = 2/5 $ Halperin state) and $p = 1$, the matrices describe the
 $ \nu = 1/2 $ spin-singlet state of the following Jain's construction:
 \begin{equation}
 \Psi_{\nu = \frac{1}{2}} = \chi_{2} \chi_{1} \chi_{1,1}.
 \end{equation}
 $ \chi_{i}, $ $ i=1,2 $, denote in  Jain's notation \cite{jain} a filled lowest
 Landau level and filled first two Landau levels without regard to the spin
 degree of freedom, and $ \chi_{1,1} $ denotes a lowest Landau level occupied
 with particles of both spins. Explicitly, the flux attaching factor,
 $ \chi_{1} \chi_{1,1} $, can be rewritten as
 \begin{equation}
\prod_{i<j} (z_{i} - z_{j})^{\frac{3}{2}} \times
\prod_{m<n} (z_{m \uparrow} - z_{n \uparrow})^{\frac{1}{2}} 
\prod_{k<l} (z_{k \downarrow} - z_{l \downarrow})^{\frac{1}{2}}  
\prod_{r,s} (z_{r \uparrow} - z_{s \downarrow})^{- \frac{1}{2}},
\end{equation}
which means that $ \frac{3}{2} $ flux quanta are attached in the usual way,
{\em i.e.}
\begin{equation}
  J_{c}(n=1,p=1)=\left( \begin{array}{cc}
              2  &  0          \\
                0     &  2          
            \end{array}   \right)
  \; \;  + \; \; 
  3   \left( \begin{array}{cc}
                                1  &  1          \\
                                1     &  1           
                               \end{array}   \right),  
 \end{equation}
 and there is also the flux attaching (without net flux attached) in the spin part
 described by $ J_{s} $ in (\ref{macs}).

By the way of the gluing theory, as in Sec. III {\bf A}, we may find
out, in this special case, that the glue vectors are
\begin{equation}
\left( \begin{array}{c}
            0 \\ 0
         \end{array} \right)_{c} \oplus
 ( 0 )_{s} \; {\rm and} \;
\left( \begin{array}{c}
            5 \\ 2
       \end{array} \right)_{c} \oplus
( 1 )_{s}.
\end{equation}
The charge part is given in the basis of the corresponding charge
excitation lattice $(L_{c}^{*})$
in which the Gram matrix is given by $(1/2) \Lambda^{-1}_{c} $ 
(see Eq.(\ref{ghcharge})) with $ m_{0} + n_{0} = 5$
and $ p = 1$. We used one-component vectors, in the spin part,
associated with the single direction that describes the spin of the
localized excitations. 
For a basis of the order-parameter lattice that is obtained by the way
of the gluing we can choose vectors:
\begin{equation}
\left( \begin{array}{c}
        4 \\ 8
       \end{array} \right)_{c} \oplus
   ( 0 )_{s},
\left( \begin{array}{c}
        5 \\ 2
       \end{array} \right)_{c} \oplus
   ( 1 )_{s}, \; {\rm and} \;
\left( \begin{array}{c}
      0 \\ 0
       \end{array} \right)_{c} \oplus
   ( 2 )_{s}
\end{equation}
(with the first and last vector describing a pure charge and pure spin
order-parameter excitation, respectively). The corresponding
three-dimensional Gram matrix is
\begin{equation}
\left( \begin{array}{ccc}
     4 & 2 & 0 \\
     2 & 3 & 1 \\
     0 & 1 & 2
       \end{array} \right).
\end{equation}

 It is straightforward to obtain higher level generalizations of this 
 construction and we will only remark that, in Jain's picture, they are
 always followed by the flux attaching procedure to the spin degrees of 
 freedom described previously.

 
 \section{Partially polarized hierarchy}
 At some level of the spin-singlet hierarchy we may allow only same-spin
 quasiparticles to make a  Laughlin state \cite{rez}. The equations that we can easily write
 down to describe this configuration are characterized by the following 
 charge matrix in the case of the first level hierarchy:
 \begin{equation}
 \Lambda_{c}=\left( \begin{array}{cc}
              m_{0} + n_{0}  &  1        \\ \label{pphcharge}
              1   &   4 p - 1          
            \end{array}   \right),
 \end{equation}
 where $ p $  is an arbitrary integer and, as before, $ m_{0} $ and $ n_{0} $ are
 of the Halperin zeroth-level state. Only in this  case,
 {\em i.e.} when $ m_{0} - n_{0} = 1 $, our charge matrix is an integer
 matrix and the invariance under the integer matrix $ S $ can be considered. If
 $ S $ is given by (\ref{transmatrix}), the transformed $ \Lambda_{c} $ is
 \begin{equation}
 J_{c}=\left( \begin{array}{cc}
              m_{0} + n_{0}  &  m_{0} + n_{0} - 1        \\ \label{jaincharge}
              m_{0} + n_{0} - 1   &   m_{0} + n_{0} - 1 + 4 p - 2         
            \end{array}   \right),
 \end{equation}
{\em i.e.} we are again in the basis of  Jain's construction for
partially polarized states \cite{jain}. When
$ m_{0} + n_{0} = 1 $ and $ p = 1 $,  the matrix also describes the simplest example
of partially polarized states with the lowest Landau level filled with both
spins and the second level (partially) filled with only one spin projection.

It is important to point out that, even in the cases of the higher
level constructions, the charge matrix is an integer
 matrix if $ m_{i}$' s and $ n_{i}$' s of the levels with no net spin polarization
 are chosen to describe a spin-singlet construction. Therefore, the presence of
  the SU(2) symmetry (which is broken in the ground state) is followed by the
   possibility to describe the invariance of the system under change of 
 the  hierarchy basis in the picture which treats the charge and
    spin degrees of freedom separately. Then the gluing theory
 description, with the integral order-parameter charge and spin
lattices  is appropriate. 
 

\section{Conclusion}
 In conclusion, the invariance of FQHE systems with spin 
 under change of the hierarchy basis can be also described in the spin-charge
 decomposition picture. We characterized the integral charge and spin
 lattices of the order parameters of the systems by stating the conditions that
 the spin-lattice Gram matrix should satisfy  to describe 
SU(2)-invariant constructions. In particular, for spin-singlet constructions, a basis
 must exist in which the Gram matrix is diagonal with $ + 2 $ or $ - 2 $
 on the diagonal. 
 This is a cosequence of the existence of as many SU(2) symmetries as
there are levels in a specific spin-singlet construction. The physical
spin, {\em i.e.} spin independent of the levels of the hierarchy, in
this basis, is a sum of the spin quantum numbers connected with each
SU(2)
symmetry.
The excitation lattices of all these systems are described
 as lattices of the gluing theory with spin only and charge only excitations
 glued in a special way. The formalism was extended to the generalized spin-singlet
 states and partially polarized states.

The lattice approach, which we used, is a systematic way to classify
all abelian FQHE states with spin. We attempted to
include all states proposed earlier in the literature, in a
description convenient for future applications. The
standard-hierarchy spin-singlet construction that we proposed, in
Sec. III {\bf B}, disagrees with the construction in Ref. \cite{rez}.
If the nominal statistics of quasiparticles is taken to be fermionic,
like in Ref. \cite{rez}, the algebraic equations of the construction
are still the same ones as in (\ref{algeq}) with $ m_{1} $ {\em even}
 (as in (\ref{algeq})).
This, although not obvious at first sight, can be seen by looking at
the wave function (of a specific spin-singlet construction) in
(\ref{longwf}),
 and reading the comments in the paragraph just following. (The
fermionic construction requires extra, flux-attaching factors (with
respect to the bosonic one), in the lowest Landau level, to be
complete.) More importantly, our construction is justified by a
clear-cut, spin-singlet condition in the same subsection.
 
 The experimental findings around filling fraction $ 3/2 $ in Ref.
 \cite{dudu} are in agreement with our conclusion that, if we do not
 consider generalized constructions, the spin-singlet states can occur
 only at the filling fractions of the form even over odd integer. 
Their results are consistent with the composite-fermion descrption of
partially polarized states \cite{jain}.  Once we have the lattice
description, presented here, we may consider the ``stability'' of
various constructions, {\em i.e.} likelihood that they correspond
to the systems with well developed plateaus in experiments \cite{fdmhald}.
It is interesting in this respect to note that the lattices
corresponding to the states described in Ref. \cite{dudu} have
positive-definite Gram matrices.

Research was supported by NSF Grant No. DMR-91-57484.
 
\end{document}